\documentclass[showpacs,preprintnumbers,pre]{revtex4}
\usepackage{graphicx}
\usepackage{dcolumn}
\usepackage{latexsym,amsmath,amssymb,graphics,color}
\usepackage{bm}
\usepackage{epsf}
\usepackage[centerlast,footnotesize,sf]{caption}
\usepackage{multirow}
\usepackage{subfigure}

\begin{document}


\title{A model for the catalytic oxidation of CO that includes CO 
desorption and 
diffusion, O repulsion, and impurities in the gas phase
}

\author{G.~M.~Buend{\'{\i}}a}\email{buendia@usb.ve}
\affiliation{Department of Physics, 
Universidad Sim{\'o}n Bol{\'{\i}}var, Caracas 1080, Venezuela}

\author{P.~A.~Rikvold}\email{prikvold@fsu.edu}
\affiliation{Department of Physics, 
Florida State University, Tallahassee, FL 32306-4350, USA}


\begin{abstract}
We present kinetic Monte Carlo simulations 
exploring the nonequilibrium phase diagram 
of a modified Ziff-Gulari-Barshad 
(ZGB) dynamic lattice-gas model for the catalytic oxidation of carbon 
monoxide (CO) on a surface.  The modified model includes the simultaneous presence of contaminants in the gas 
phase, CO desorption, CO diffusion, 
and strong repulsion between adsorbed oxygen (O) atoms; all of which have been observed in 
experimental systems. We find that the strong O-O repulsion produces 
higher reaction rates, albeit in a reduced reactive pressure window. 
In systems with impurities, the 
CO$_2$ production rate is greatly reduced, but this effect is mitigated 
by CO desorption and diffusion. 
CO desorption has the effect of widening the reactive pressure window, while 
CO diffusion has the effect of increasing the reaction rate. 
In some parameter regimes the presence of impurities destroys the discontinuous 
transition between the reactive and high CO coverage phases. 

\end{abstract}

\date{\today}


\pacs{82.65.+r, 64.60.Ht, 82.20.Wt}

\maketitle

\section{Introduction}
\label{sec:I}

The catalytic oxidation of CO on a surface is perhaps the most studied example 
of  heterogeneous catalysis. Besides its obvious industrial applications, this 
reaction constitutes a prime example of a nonequilibrium system that exhibits a 
rich variety of behavioral patterns and complex irreversible critical behaviors 
\cite{marro99, chri94}.  In recent years, studies based on complex  
lattice-gas models, combined with detailed molecular information and realistic input energetics 
obtained from density functional theory in combination with experiments,
have provided detailed results for some aspects of this 
reaction~\cite{volk01,petr05,liu06,naga08,roga08, liu09, hess12}. 
For a recent, extensive review, see~\cite{liu13}.

However, a comprehensive understanding of heterogeneous catalysis 
and the associated, nonequilibrium phase diagrams 
remains a challenging problem, and there is still much to be learned from simple models that 
can easily be manipulated to incorporate 
different physical features. The well-known Ziff-Gulari-Barshad (ZGB) 
dynamic lattice-gas model~\cite{ziff86}
is an example of such a simple model that has proved to be a 
fruitful laboratory for testing the effects of various processes on catalytic reactions,  
and for exploring out-of-equilibrium phase transitions in general~\cite{losc03}. 
The model has recently been used to test novel algorithms 
for studying rare events \cite{adam10}.

An important 
characteristic of this model is that it can be enriched by the inclusion of 
different physical processes without losing its 
intrinsic simplicity. In the original ZGB model, the CO-O$_2$ 
(or more generally the A-B$_2$) reaction occurs via a 
Langmuir-Hinshelwood mechanism that involves only one parameter, the relative 
partial pressure of CO in the gas-phase. The model exhibits 
two phase transitions: a continuous one at low CO pressure, between an oxygen 
poisoned phase and a reactive phase, and a discontinuous one, at higher CO 
pressure,  between the reactive phase and a CO-poisoned phase. These features 
make the model a useful tool to explore the nature of 
transitions in nonequilibrium systems \cite{grin89, tome93, mon00, mach05b}.  
The continuous transition at low CO pressure 
(known to belong to the the directed percolation universality class \cite{grin89})
has not been experimentally observed, but it can be easily eliminated by 
making minor modifications to the model \cite{wint96, jame99,buen09}, as 
discussed below. 

With the purpose of 
approaching the CO-O$_2$ reaction in a somewhat more realistic way, 
and to understand the effects of different processes on 
its nonequilibrium phase diagram, 
our aim in this paper is to 
subject the ZGB model to the simultaneous influence of 
several perturbing processes, so that their interactions can be 
directly ascertained. 
In previous work we have studied the effects of some of 
these processes individually or in smaller combinations  
\cite{buen09,buen12,buen13}.  
These include the temperature-related 
effect of CO desorption \cite{kauk89, bros92, alba92,buen09}
and a modification that mimics experimental results that 
indicate that the oxygen atoms suffer a strong mutual repulsion once on the surface 
\cite{wint96, jame99}.  
Several of these modifications have been studied separately 
and in various combinations \cite{pav01,hua02,buen06}. 
In order to better understand the effects of pollutants, always 
present in real environments, we also incorporated impurities in the gas phase 
\cite{buen12,buen13, bust00, hua03}. 
Under realistic industrial conditions the temperature and 
pressure can reach high values, regimes in which the validity of this 
simple model is limited. Nevertheless, one can still gain 
qualitative insight into the 
effects of impurities on the process and, most importantly, how they affect 
the nature of the nonequilibrium phase transitions. 
The results show that the impurities significantly affect the efficiency of the 
process ~\cite{volk01, buen12,buen13,BZOV09,sinh10, lore02}. One interesting 
effect of adding impurities is the disappearance for low impurity desorption rate 
of the first-order phase 
transition between the reactive phase and the CO-poisoned phase \cite{buen12,buen13, bust00}. 
This result has been further 
confirmed for the case of quenched defects by a recent study that 
indicates that first-order transitions do not 
exist in nonequilibrium disordered systems with absorbing states \cite{vill14}.

CO desorption prevents the formation of a CO-poisoned absorbing state, and 
consequently the abrupt transition from the reactive state to the low reactive 
state becomes reversible 
\cite{losc03,tome93,mon00,mach05b,kauk89,bros92,alba92}. 
A totally CO-poisoned state cannot 
be  achieved experimentally due to the nonvanishing CO desorption. The 
first-order nature of the transition only remains 
up to a critical value of the CO 
desorption rate \cite {losc03,buen12,alba92}, where it terminates at a critical point  
that belongs to the two-dimensional Ising universality class \cite{tome93}. 
Our previous 
studies indicate that CO desorption can counteract the negative effect of 
the impurities by widening the region where the system remains catalytically 
active \cite{buen13}.   

In the present work we are particularly interested in studying 
how CO diffusion, when added to the other effects (O-O repulsion, CO desorption, and 
gas-phase impurities), alters the behavior of the system. 
It is well known that the 
mobility of the adsorbates plays an important role in the overall reactivity of 
the surface. Kinetic Monte Carlo studies indicate that CO diffusion is the 
key to the qualitative differences between the CO electrooxidation dynamics on 
rhodium and platinum \cite{hous07}, and that it synchronizes oscillations in models that include 
surface reconstruction \cite{gelt98}. 


Real diffusion 
is usually very fast, especially in systems at low pressure, with very large diffusion 
lengths that determine the characteristic length scale of spatial patterns and 
the propagation of reaction fronts \cite{tamm98, liu00}. The 
wide range of time scales makes computer simulation of this rapid 
diffusion very challenging. Several approaches have been proposed, from 
coarse-grained methods,  generalized Metropolis simulations, even hydrodynamics 
based algorithms \cite{liu13}. 
An  interesting approach is to treat the diffusion by a mean 
field and the reaction with Monte Carlo \cite{tamm95}. 
In the present work we treat diffusion as a relatively slow process. Even in this limit, diffusion 
proves to have significant effects on the nonequilibrium 
phase diagram of the model. 

The rest of this paper is organized as follows. In Sec.~\ref{sec-Mod} we describe in 
detail how the ZGB model is modified to eliminate the unphysical continuous phase 
transition at low CO pressure by including strong repulsion between  
adsorbed O atoms, 
how we incorporate CO desorption and add diffusion to the processes included in our previous studies, 
and how we take into 
account  the impurities in the gas phase. In Sec.~\ref{sec-Res} 
we present our numerical 
results for the modified model, and in Sec.~\ref{sec-Con} we present our conclusions.

\section{Model and Simulations}
\label{sec-Mod}
 In this work we study the catalytic oxidation of CO on a surface immersed in a 
gas phase that consists of a mixture of CO, O$_2$, and inert impurities, X, 
in different proportions. As in the original ZGB model, CO and O$_2$ can be 
adsorbed on the surface, and once there they react to produce CO$_2$ that is 
immediately desorbed. The impurities can be adsorbed or desorbed at single lattice sites on 
the surface, where they do not react with the other adsorbates. (Several common catalytic poisons, 
such as sulfur and lead, adsorb at single lattice sites, although the chemistry of the 
reaction inhibition can be more complicated than the simple site blocking we consider here.) 
This model differs from the original ZGB model in the presence of 
impurities in the gas phase, in the entrance mechanism for the O atoms, and in 
the existence of CO desorption and diffusion. The modification of the entrance 
mechanism is 
inspired by experimental results that indicate that the O atoms tend to repel 
each other once on the surface ~\cite{wint96, alba94}. 
Recently, quite sophisticated models amenable to analytic treatment have been 
introduced  for the dissociative adsorption of  O$_2$.  Some of these 
also incorporate exclusion of nn pairs of adsorbed O, and successfully 
reproduce experimental results that describe the behavior of the sticking coefficient 
versus the coverage \cite{liu13}.

We impose the condition that an  O$_2$ molecule can 
be adsorbed on two next-nearest-neighbor (nnn) vacant sites (separated by 
$\sqrt2$ lattice constants) only if the six nearest neighbors (nn) to 
these sites contain no O. In the original ZGB model the O atoms are adsorbed on 
nn empty sites. Thus, in the modified model, the adsorption of an O$_2$ 
molecule requires the existence of eight sites that do not contain O. Clearly, this 
requirement eliminates the unphysical O-poisoned phase that appears in the 
standard ZGB model. In the literature this 
adsorption prescription is known as the eight-site rule~\cite{brun84, chan87,jame99}.  
Previous work shows that just 
requiring that the O atoms enter nnn sites is sufficient to ensure that 
the O-poisoned phase disappears \cite{buen12, buen13, ojed11}. 
The mechanism is essentially that,
at low values of $y$ the surface is mostly covered by
O. The few isolated empty sites can only be filled with
CO that will react with one of its nn O and leave two nn empty
sites. In the original ZGB model these empty nn sites would,
with very high probability, be filled by O, eventually
poisoning the surface. With the requirement of nnn O adsorption, they can be filled only
with CO that will continue reacting with their O neighbors, preventing the 
buildup of a complete O coverage \cite{buen12}.

We note that next-nearest-neighbor O adsorption is not the only mechanism that can 
remove the unphysical O-poisoned phase. Another possibility is an Eley-Rideal 
mechanism that allows reaction between CO molecules in the gas phase and adsorbed 
O atoms \cite{buen09}. The incorporation of CO desorption and 
diffusion has the purpose to emulate temperature effects that are known to be very important in 
the catalytic process. In contrast, O atoms are relatively immobile at temperatures 
below 400~K \cite{ehsa89}, and in general O typically has a significantly higher diffusion 
barrier than CO \cite{liu13,RENI99,MITS05}. We therefore choose to neglect O diffusion 
in our model. 
The impurities in the gas-phase are added to study 
the system under somewhat more realistic industrial conditions.
  
The model  is simulated on a square lattice of
linear size $L$ that represents the catalytic surface.  A Monte Carlo
simulation generates a sequence of trials: CO, X, or O$_2$ adsorption, CO or 
X desorption, or CO diffusion.  A site $i$ is selected at random.  If it is occupied by 
CO, we attempt desorption with probability $k_{\text{co}}$ and  diffusion with 
probability $d$.  ($k_\text{co}+d \le 1$.) For the diffusion step we randomly 
choose one of the nn of $i$. If it is empty, we move the CO to the new 
site.  If $i$ is occupied by an X, we attempt desorption with probability 
$k_{\rm x}$. 
If $i$ is empty, we
attempt adsorption: CO with probability $y$,  X with probability $y_{\rm x}$, or  
O$_2$ with probability $1-y-y_{\rm x }$. These probabilities are the relative 
impingement rates of the molecules and are proportional to the amounts of the 
different species in the gas phase.   The dissociative adsorption of an O$_2$ 
molecule 
requires the existence of a pair of vacant nnn sites and that the six nn to the 
pair do not contain an O atom. A nnn of site $i$ is selected at random; if 
it is occupied the trial ends, if not, the six nn of the pair are checked, if none 
of them contains an O molecule, the adsorption proceeds and the O$_2$ molecule 
is adsorbed and dissociates into two O atoms. After CO adsorption or 
diffusion or O$_2$ adsorption, all nn pairs are checked in 
random order. Pairs consisting of a nn CO and O 
react: a CO$_2$ molecule is released, and two nn sites are vacated. 
A schematic representation of this algorithm is given by the equations,
\begin{eqnarray}
\text{CO(g)} + \text{S} & \rightarrow & \text{CO(a)}
\nonumber \\
\text{O}_2 + 2\text{S} & \rightarrow & 2\text{O(a)}
\nonumber \\
\text{CO(a)} + \text{O(a)} & \rightarrow & \text{CO}_2\text{(g)} + 2\text{S}
\\
\text{X(g)} + \text{S} & \rightarrow & \text{X(a)}
 \nonumber \\
\text{X(a)} &\rightarrow& \text{X(g)}+ \text{S}
\nonumber \\
\text{CO(a)} &\rightarrow& \text{CO(g)}+ \text{S}
\nonumber\\
\text{CO(a)}_{\text{S}}+\text{S'} &\rightarrow& \text{CO(a)}_{\text{S'}}+\text{S}
\;
\nonumber
\end{eqnarray}
Here, S and $\text{S'}$ represent  empty sites on the surface, $\rm g$ means gas 
phase, and $\rm a$ means adsorbed. CO(a)$_{\text{S}}$ means an adsorbed CO at 
the S site.
The first three steps correspond to a Langmuir-Hinshelwood mechanism. The fourth and fifth 
terms represent the adsorption and desorption of an impurity, respectively. 
The sixth term represents CO desorption, and the seventh represents diffusion 
of a CO molecule from site S to its nearest neighbor site S'. 

In our simulations the surface is represented by a lattice of size 
120$\times$120 with periodic boundary conditions. 
(Additional results for a 200$\times$200 system are included in Fig.~\ref{f5}.) 
The time unit is one Monte Carlo Step per Site (MCSS), during which each
site is visited once on average. Averages are taken over $10^5$
MCSS after $10^5$ MCSS that are used to achieve a stationary state. 

Coverage is defined in the usual way as the fraction of sites on the surface 
occupied by an adsorbate. We calculate the CO, O, and X coverages and the rate 
of production of CO$_2$.  For each value of $y$ and $y_{\rm{x}}$ we start from an 
empty lattice and let the system reach a steady state before taking 
measurements.

\section{Results} 
\label{sec-Res}

To understand the effects of the strong 
repulsion between adsorbed O and the presence of CO diffusion and desorption, 
in Sec.~\ref{sec-ResA} we first 
present the case in which there are no impurities  
and compare the results with those obtained by a previous model 
~\cite{buen13} that does not 
include CO diffusion or the new entrance mechanism for O. 
Then, in Sec.~\ref{sec-ResB}, we present the results for the full model 
described in Sec.~\ref{sec-Mod}, which includes  impurities.

\subsection{ Without  impurities, $y_{\rm x}=0$}
\label{sec-ResA}

In this subsection we introduce CO 
diffusion in two models that both include CO desorption, but differ in the 
entrance mechanism for oxygen.  In 
one model, that for simplicity will be labeled A in this subsection, the O atoms enter a pair 
of nnn sites without any other 
requirement ~\cite{buen13}. In the second model (in this subsection labeled B), the entrance of the O 
atoms at a pair of vacant nnn sites occurs 
only if the six nn of the pair are free of oxygen.  
Thus, the dissociative adsorption of O$_2$ in model A requires two empty sites, while 
in model B it requires two empty sites and six additional sites free of O. In this subsection, 
neither model includes impurities.

\begin{figure}
\includegraphics[scale=.30]{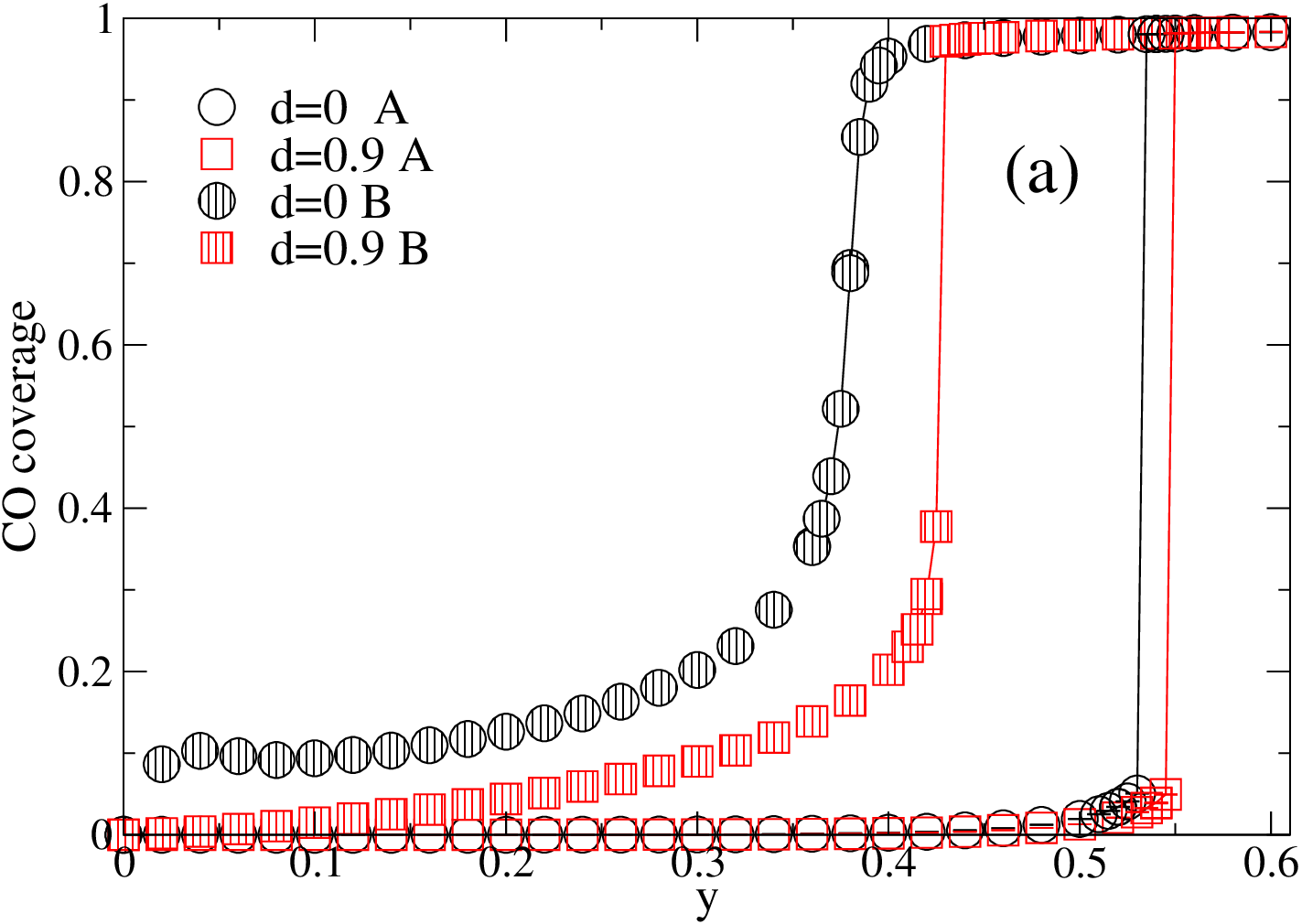}
\hspace{0.2truecm}
\includegraphics[scale=.30]{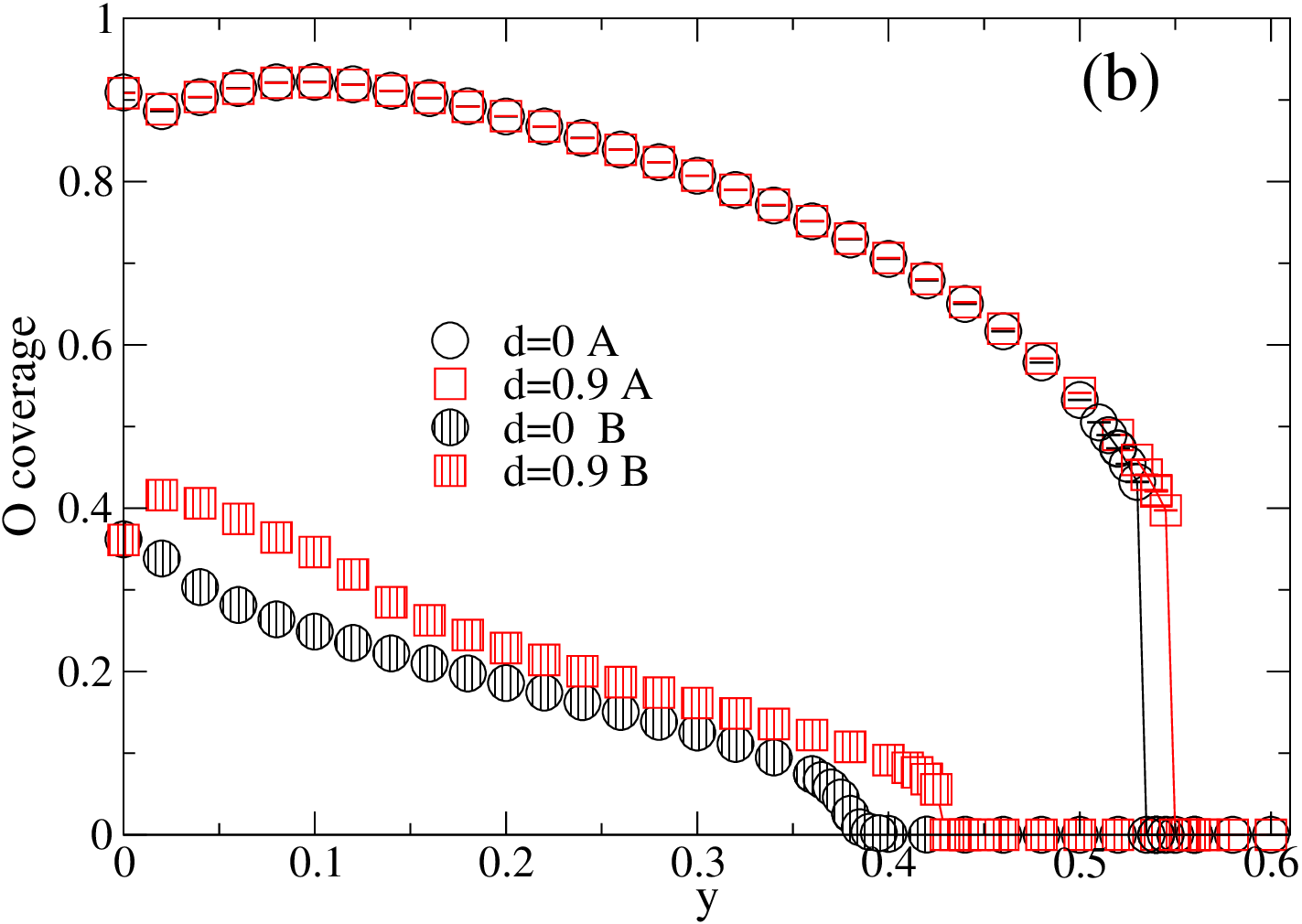}
\vskip 1.5truecm 
\includegraphics[scale=.30]{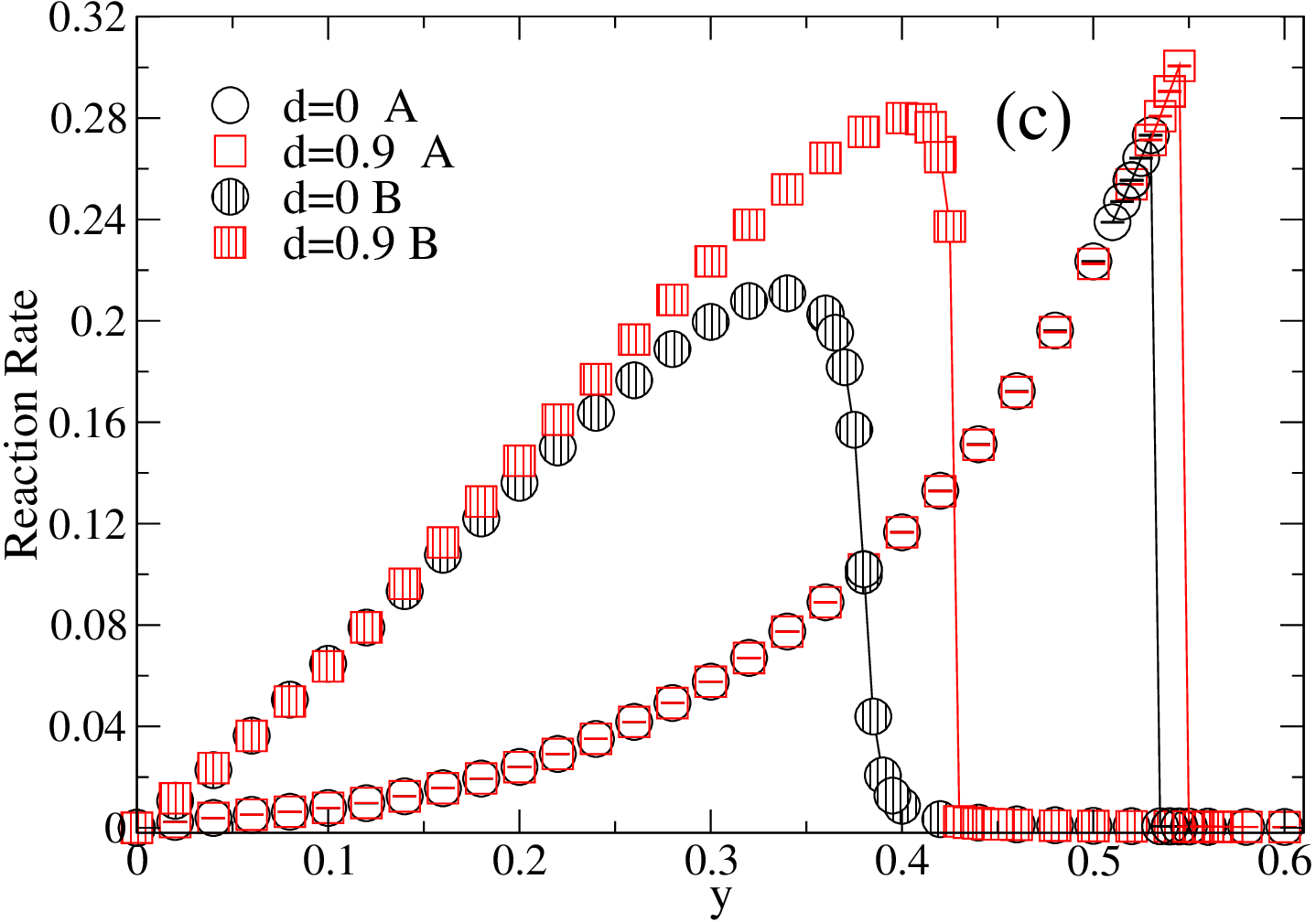}
\caption[]{
Coverages and reaction rate vs $y$ in the case that there are no impurities 
($y_{\rm x}=0$) 
for models A (empty symbols) and B (filled symbols) with the values of the diffusion probability 
$d$ indicated in the figures. (a) CO coverage, (b) O coverage, (c)  CO$_2$ 
production rate.   $k_{\rm co}=0.01$, below the critical value for the original ZGB model with CO desorption.
The results in this and subsequent figures are for systems of size 120$\times$120 lattice sites.}
\label{f1}
\end{figure}

\begin{figure}
\includegraphics[scale=.30]{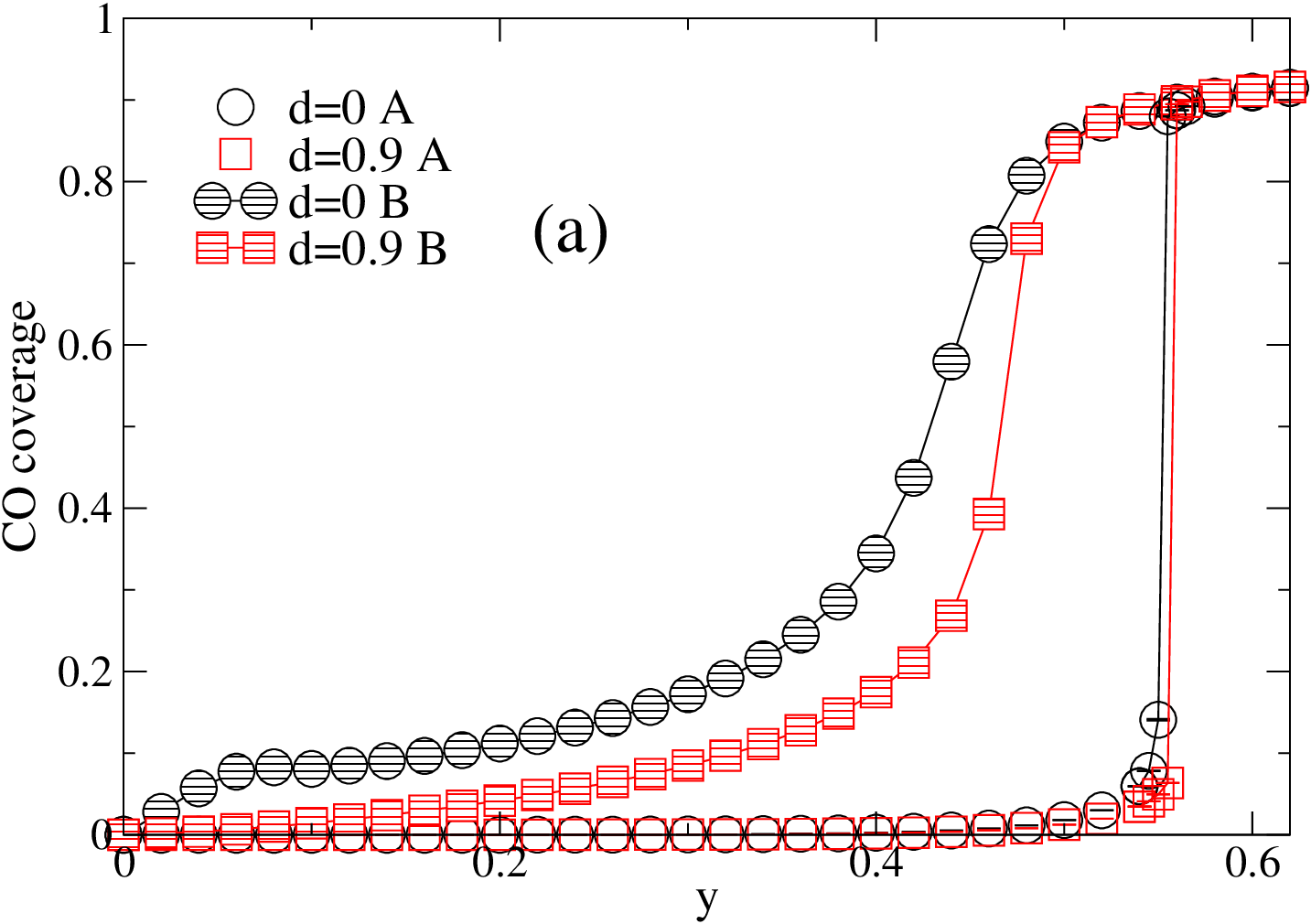}
\hspace{0.3truecm}
\includegraphics[scale=.30]{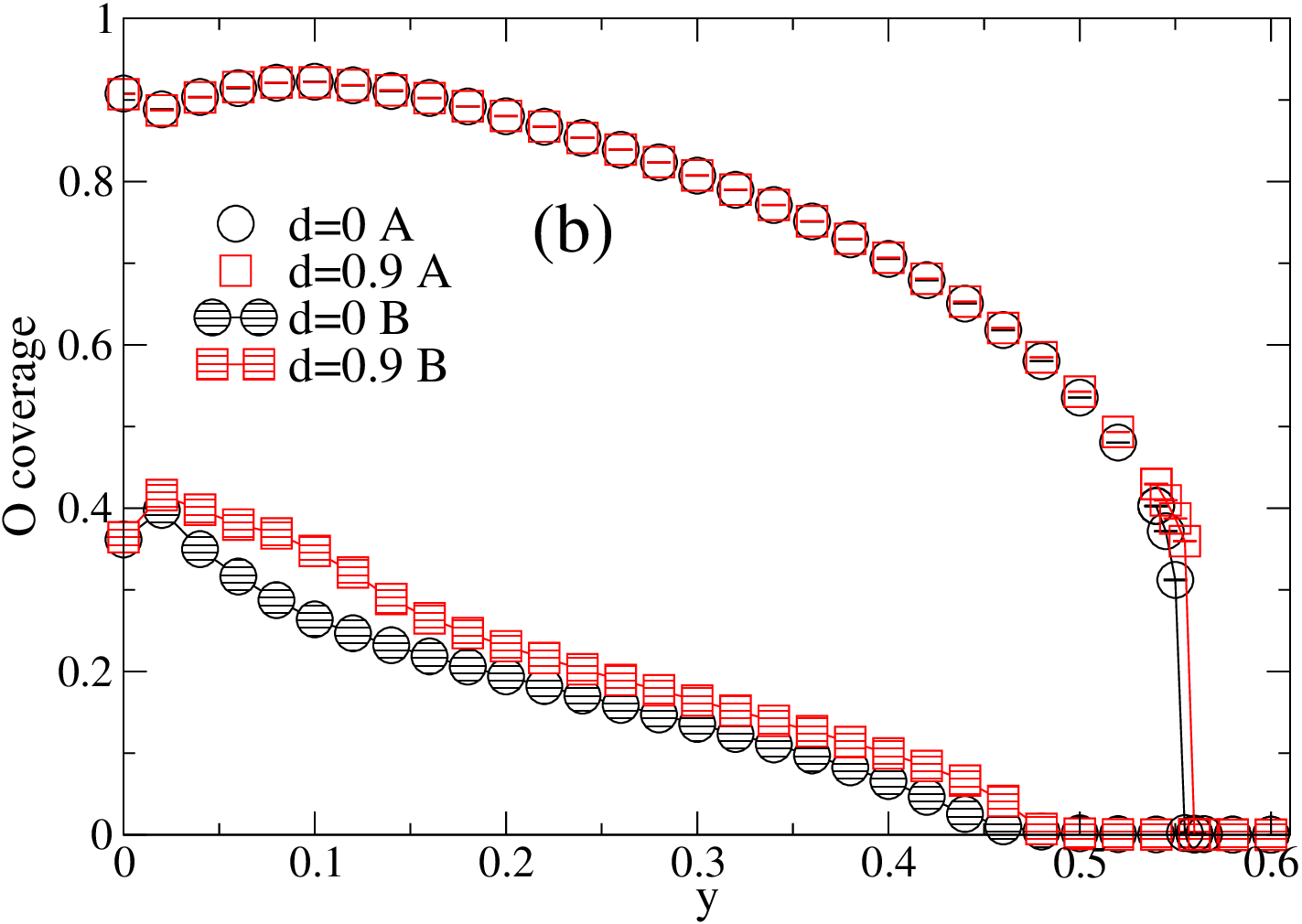}
\vskip 1.5truecm
\includegraphics[scale=.30]{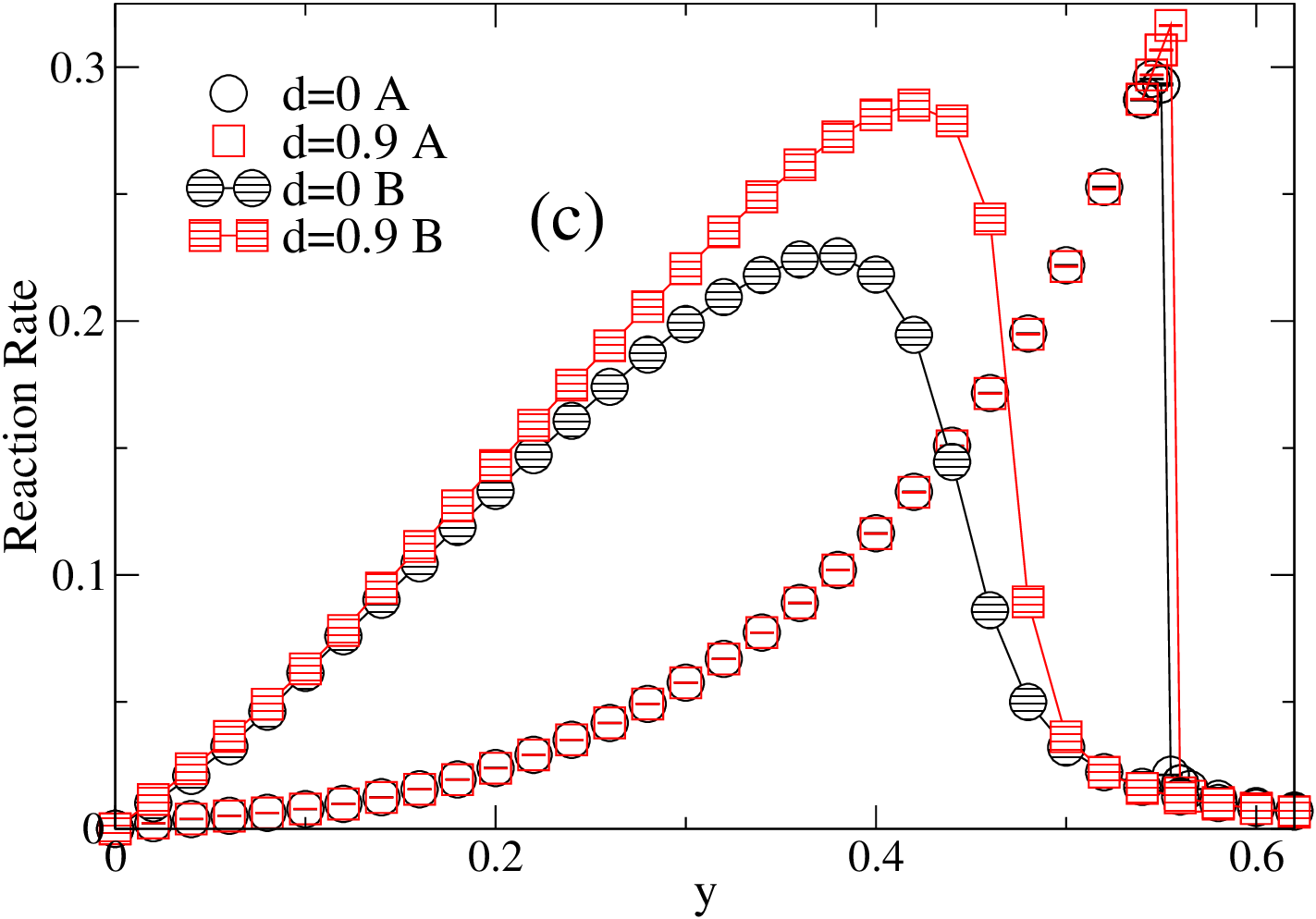}
\caption[]{
Coverages and reaction rate vs $y$ in the case that there are no impurities ($y_x=0$) 
for models A and B with the values of the diffusion probability $d$ indicated in the figures. (a) 
CO coverage, (b) O coverage, (c)  CO$_2$ production rate.   $k_{\rm co}=0.05$, above the critical value for the 
original ZGB model with CO desorption.
}
\label{f2}
\end{figure}

In Fig.~\ref{f1} and Fig.~\ref{f2} we compare the coverages and reaction 
rates of the two models, for a 
small CO desorption rate, $k_{\rm{co}}=0.01$, and for a larger one, 
$k_{\rm{co}}=0.05$, respectively. In each case we plot the results for the case 
without CO diffusion ($d=0$), and with a high diffusion rate ($d=0.9$). The 
first effect associated with the new entrance mechanism for oxygen, is that 
the transition to the CO poisoned state is smoother for model B, and that it 
occurs at a lower value of $y$. The latter effect is due to the higher 
CO coverage in the active phase in model B, as observed in Fig.~\ref{f1}(a) and 
Fig.~\ref{f2}(a). As expected, the O coverage, Fig.~\ref{f1}(b) and 
Fig~\ref{f2}(b), is much smaller in model B. It is well 
known that the O poisoned phase that characterizes the standard ZGB model is 
not present in these models, in which the O enter at nnn 
sites ~\cite{buen12,buen13,ojed11}. 
As expected, the coverage of empty sites, which can be trivially 
calculated from the CO and O coverages, is much larger in model B.  The reaction 
rate behaves quite differently in the two models, Fig.~\ref{f1}(c) and Fig.~\ref{f2}(c). In model A it 
starts growing very slowly but increases quite rapidly as the transition point 
$y_c$ is 
reached. In model B it increases at an almost constant rate until it 
reaches its maximum. As a result, below
the transition point (which depends on the model), for the same value of $y$, 
the reaction rate is much higher in model B.  As 
has also been observed in other models, the reaction rate increases with  
$k_{\rm{co}}$. It is generally expected that diffusion  
increases the reaction rates, since the mobility of the adsorbates facilitates 
the number of encounters between the reactants. However this effect is only 
relevant when the lattice has a significant number of empty sites that 
allow diffusion to take place, and a high coverage of the diffusing species. 
In the reactive region model B has a larger number of empty 
sites and also a 
larger CO coverage. This explains why the reaction rates, Fig.~\ref{f1}(c) and Fig.~\ref{f2}(c), 
show a marked increase with $d$. In contrast, in model A the CO 
coverage and the number of empty sites below the transition point is very small. 
Therefore, the diffusion effects are almost negligible. These 
results indicate that the system behaves quite differently when a strong 
repulsion effect between the O atoms is taken into account, and in this 
case the mobility of the CO is quite relevant. Notice in Fig.~2 that, when 
$d=0$, model A has a sharp transition to the CO rich phase, even for $k_{\rm 
co}=0.05$, larger than $k_{\rm{co}}^{crit} \lesssim 0.04$ for the 
original ZGB model with CO desorption ~\cite{tome93,mach05b, bros92,chan14}. This is 
consistent with previous results ~\cite{ojed11} that indicate that the critical value of $k_{\rm{co}}$ 
is increased in model A.
On the other hand, in model B, which includes the eight-site rule, it appears that the  critical value of $k_{\rm{co}}$ is 
reduced relative to the ZGB model with CO desorption, leading to the smooth maxima in the reaction rate. 

Diffusion increases the spatial reaction range, leading to an increase in the critical value of $k_{\rm{co}}$
\cite{chan14,48}. 
In Fig.~\ref{f1} this has the effect that the transition in Model B is continuous for $d=0$, but discontinuous 
for $d=0.9$. At the higher value of $k_{\rm{co}}$ shown in Fig.~\ref{f2} however, 
the transition is continuous both with and 
without diffusion. In Model A, the size of the discontinuity is seen to increase with $d$ in both Fig.~\ref{f1} and 
Fig.~\ref{f2}. 

\subsection{With impurities, $y_x > 0$}
\label{sec-ResB}

\begin{figure}
\includegraphics[scale=.3]{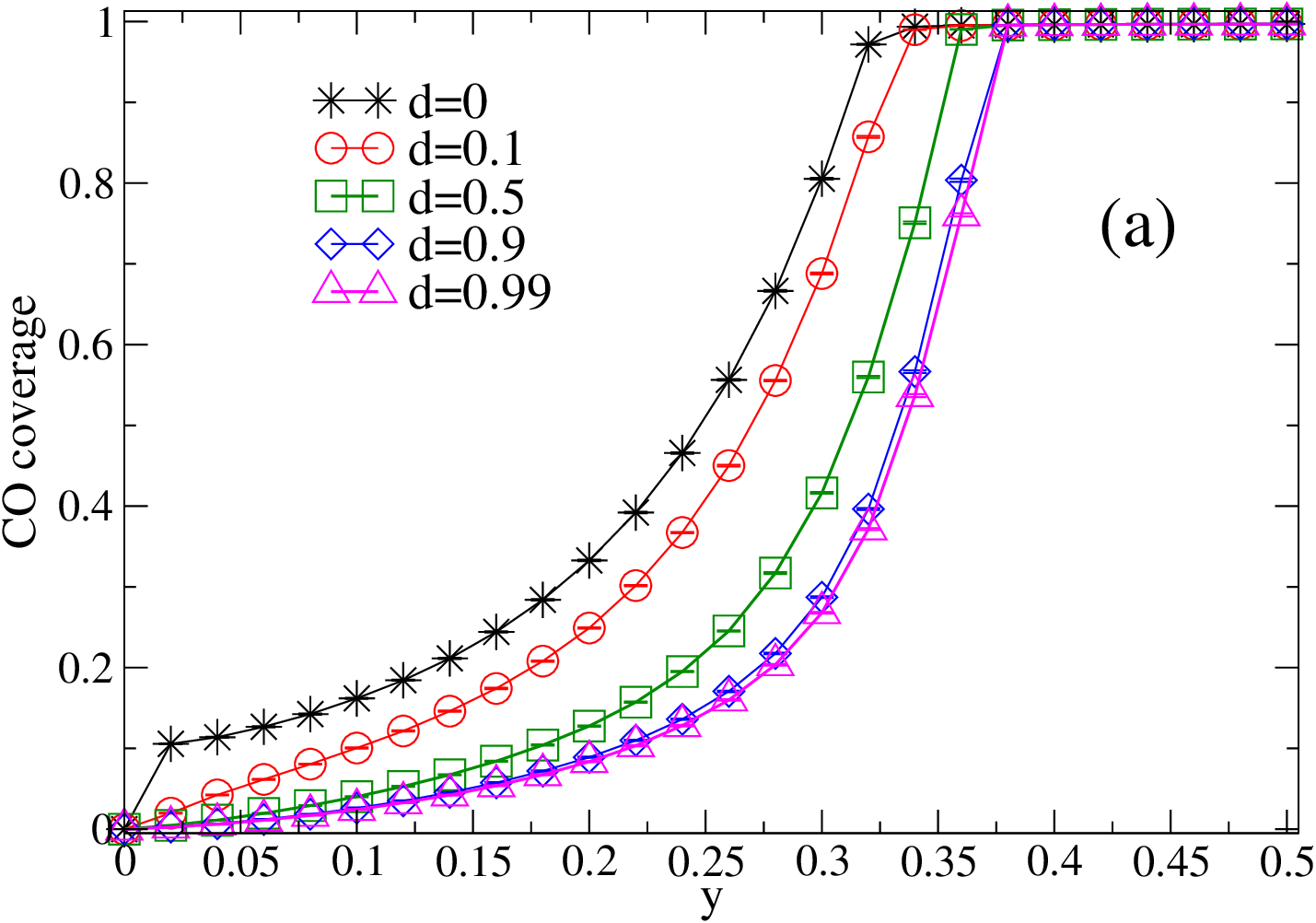}
\hspace{0.2truecm}
\includegraphics[scale=.3]{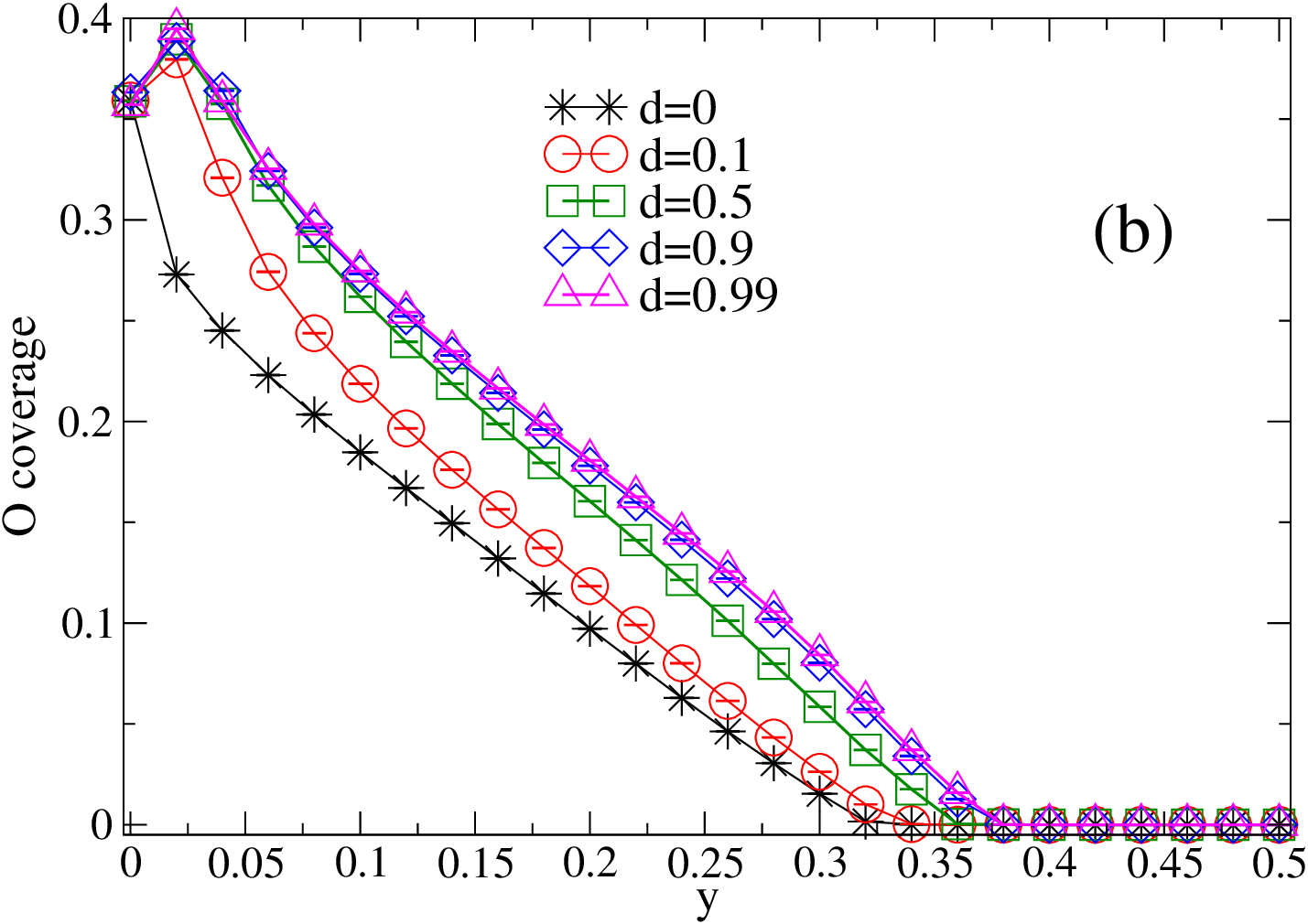}
\vskip 1.0truecm
\includegraphics[scale=.3] {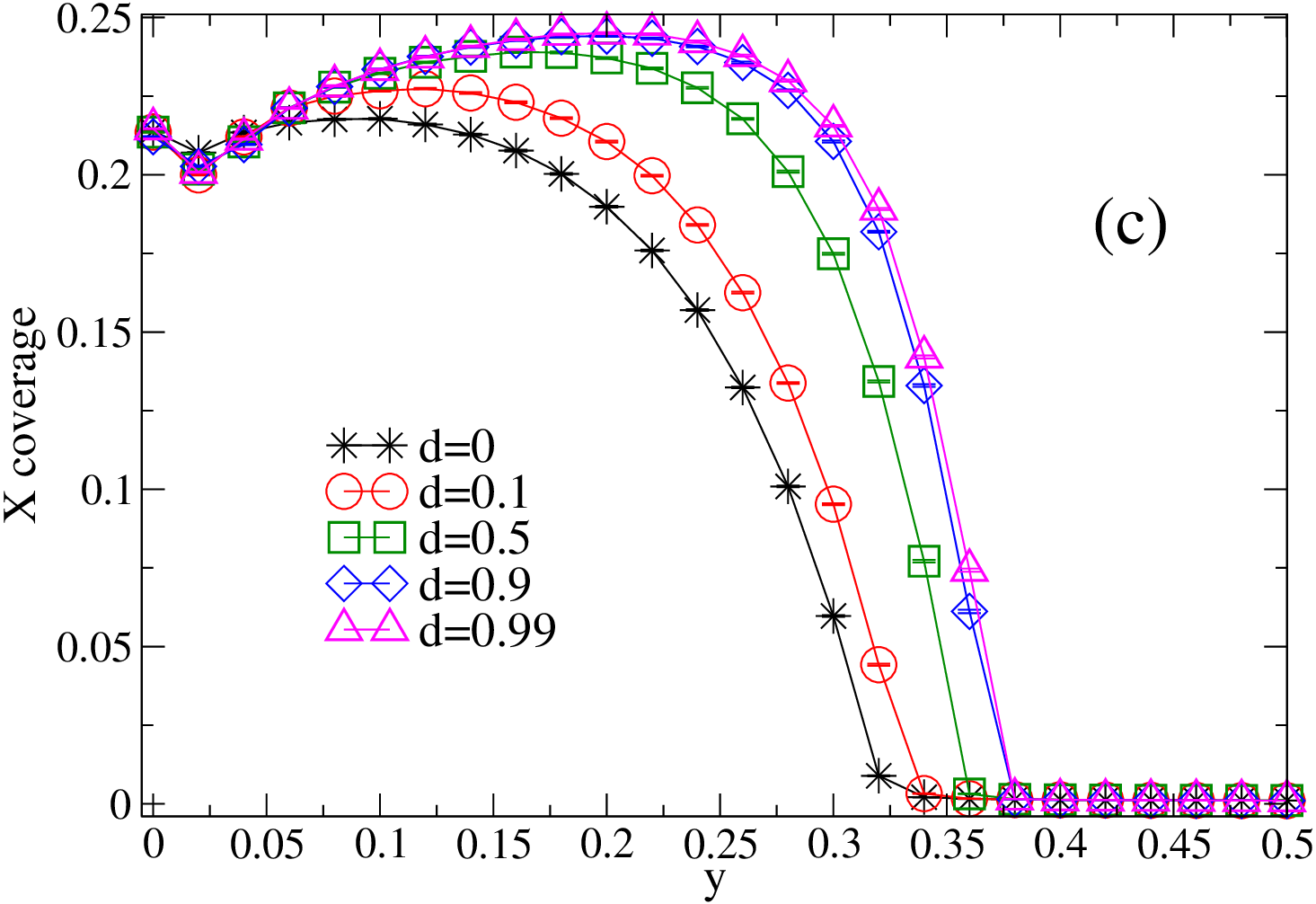}
\hspace{0.2truecm}
\includegraphics[scale=.3]{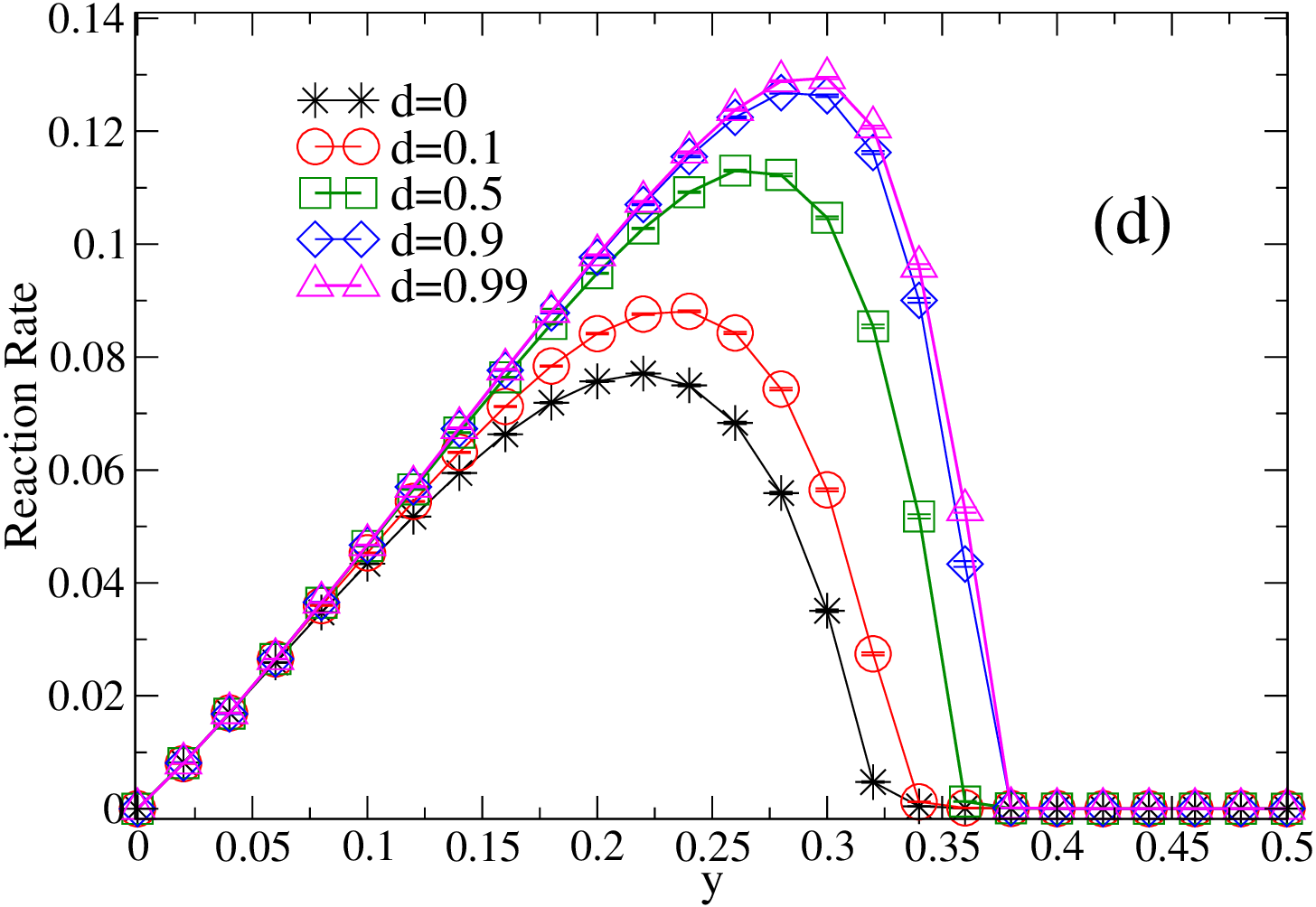}
\caption[]{
Coverages and reaction rate vs $y$ when the partial pressure of impurities is $y_{\rm 
x}=0.005$,  (model B) with the values of $d$ indicated in 
the figures. (a) CO coverage, 
(b) O coverage, (c) X coverage and (d)  CO$_2$ production rate.   
$k_{\rm{co}}=0.001$ and $k_{\rm{x}}=0.01$.}
\label{f3}
\end{figure}

\begin{figure}
\includegraphics[scale=.35]{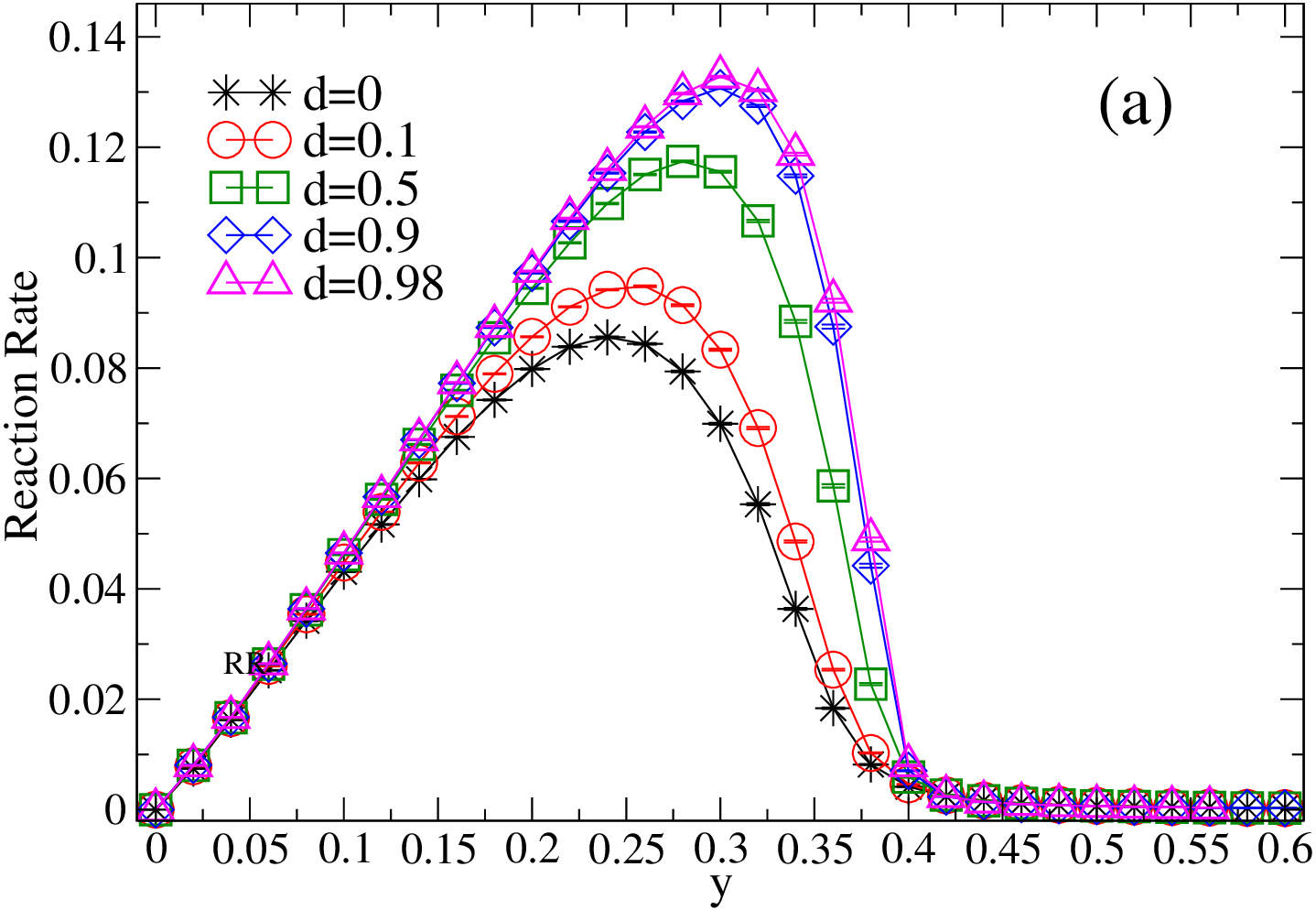}\\
\vskip 1.0truecm
\includegraphics[scale=.35]{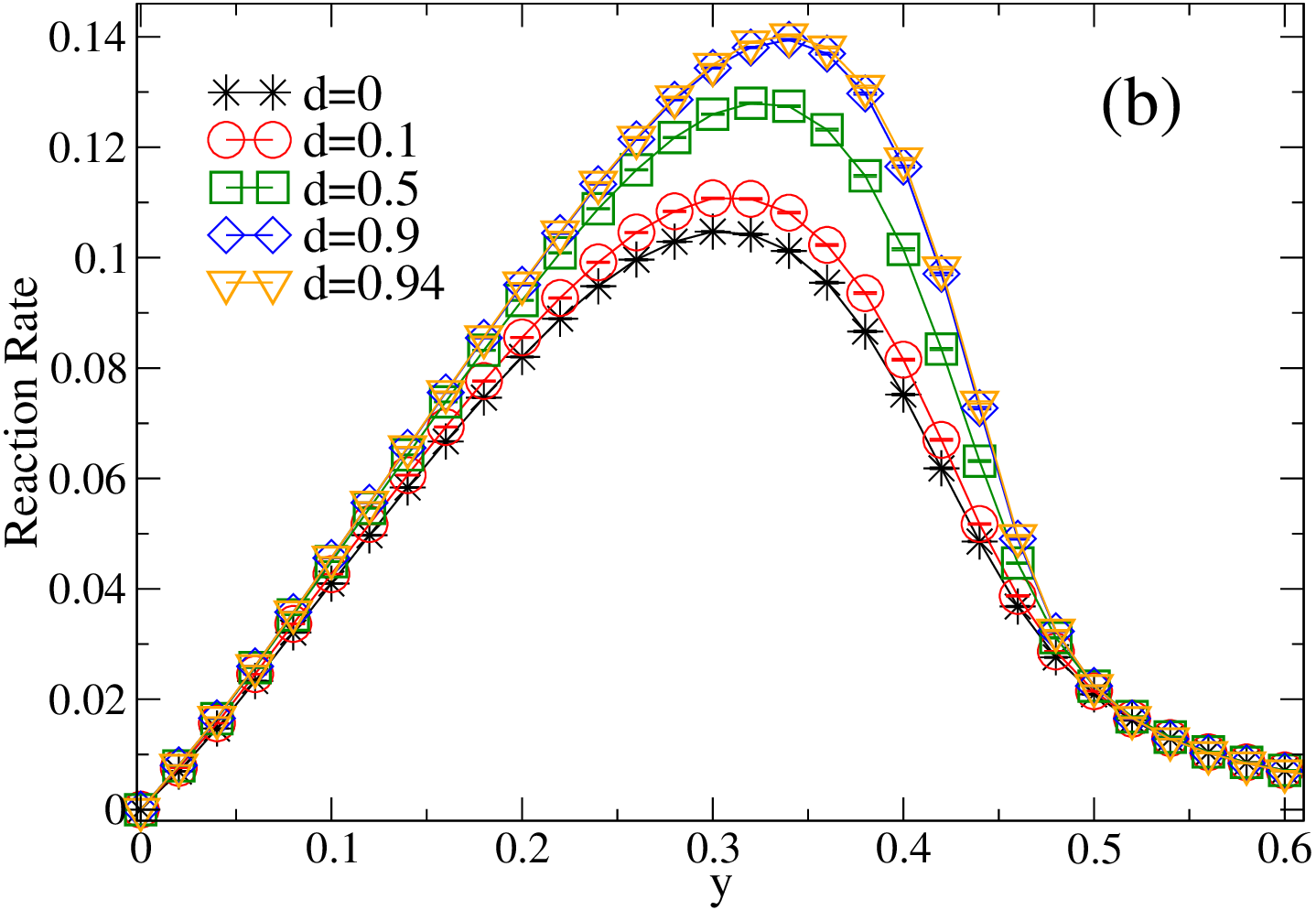}
\caption[]{
Reaction rates vs $y$ when the pressure of impurities is 
$y_{x}=0.005$ (model B), at  $k_{\rm x}=0.01$ with the values of $d$ 
indicated in the 
figures. (a) $k_{\rm co}=0.01$. (b) $k_{\rm co}=0.05$. Notice that for the latter value of $k_{\rm co}$, 
the phase transition is replaced by a smooth crossover.}
\label{f4}
\end{figure}

In this subsection we study the effects of impurities in the gas phase, considering only the 
model labeled in the previous subsection as B, which includes the new 
mechanism for the entrance of oxygen, as well as CO diffusion and desorption. 
Since this 
is the only model we consider in this subsection, we drop the label B. The 
effects of 
impurities in the gas phase in the model without CO diffusion and without the 
strong repulsion between adsorbed O, were discussed 
in Ref ~\cite{buen13}. The impurities can be desorbed from the surface with 
probability $k_{\rm x }> 0$. Previous studies ~\cite{buen12, buen13} indicate 
that 
when non-desorbing impurities,  $k_{\rm x}=0$, are present, the steady state of 
the system has reaction rate zero. We verified that this is also the case 
in the model described here; the existence of CO diffusion does not alter this 
behavior. We fix the 
partial pressure of impurities to $y_{\rm x}=0.005$. To analyze the 
effects of $k_{\rm co}$, $d$, and $k_{\rm x}$, 
we start by fixing $k_{\rm x}$ and study the dependence of the 
system on $d$ 
for different values of $k_{\rm{co}}$.  As an example, in Fig.~\ref{f3} we present the 
coverages and the reaction rates with $k_{\rm co}=0.001$ and 
$k_{\rm x}=0.01$ for 
different values of $d$.  As $d$ increases, the CO coverage 
decreases and the O and X coverages increase. The reaction 
rate increases considerably with $d$. Notice that when $d$ reaches a limiting 
value of about 0.9, further increase does not affect the coverages or the reaction rate.  
This can be seen in Fig.~\ref{f4}, where we plot the reaction rate vs 
$y$ for several values of $d$ for $k_{\rm co}=0.01$ (a) and 
$k_{\rm{co}}=0.05$ 
(b). Since, for the same value of $y$, the CO coverage becomes smaller as $d$ 
increases  (see Fig.~\ref{f3}(a))  and the number of empty sites 
remains almost the same above a certain value of $d$, it is easy to see why 
increasing $d$ beyond a certain point does not have any 
effect on the behavior of the system. In our algorithm the maximum value that $d$ 
can take is $1-k_{\rm{co}}$. Comparing Fig.~\ref{f1}(c) and Fig.~\ref{f2}(c) with Fig.~\ref{f4}(a) and 
Fig.~\ref{f4}(b), respectively, we see that the presence of impurities substantially 
reduces the reaction rate of the system. Comparing, Fig.~\ref{f3}(d) with Fig.~\ref{f4} we see 
that, as expected, the reaction rate increases when $k_{\rm{co}}$ increases, 
however this effect seems to be only significant when $d$ 
is small. Comparing Fig.~\ref{f4}(a) and Fig.~\ref{f4}(b), it is evident that the most relevant 
effect is that the reactive window increases considerably when $k_{\rm co}$ 
increases.

\begin{figure}
\includegraphics[scale=.35]{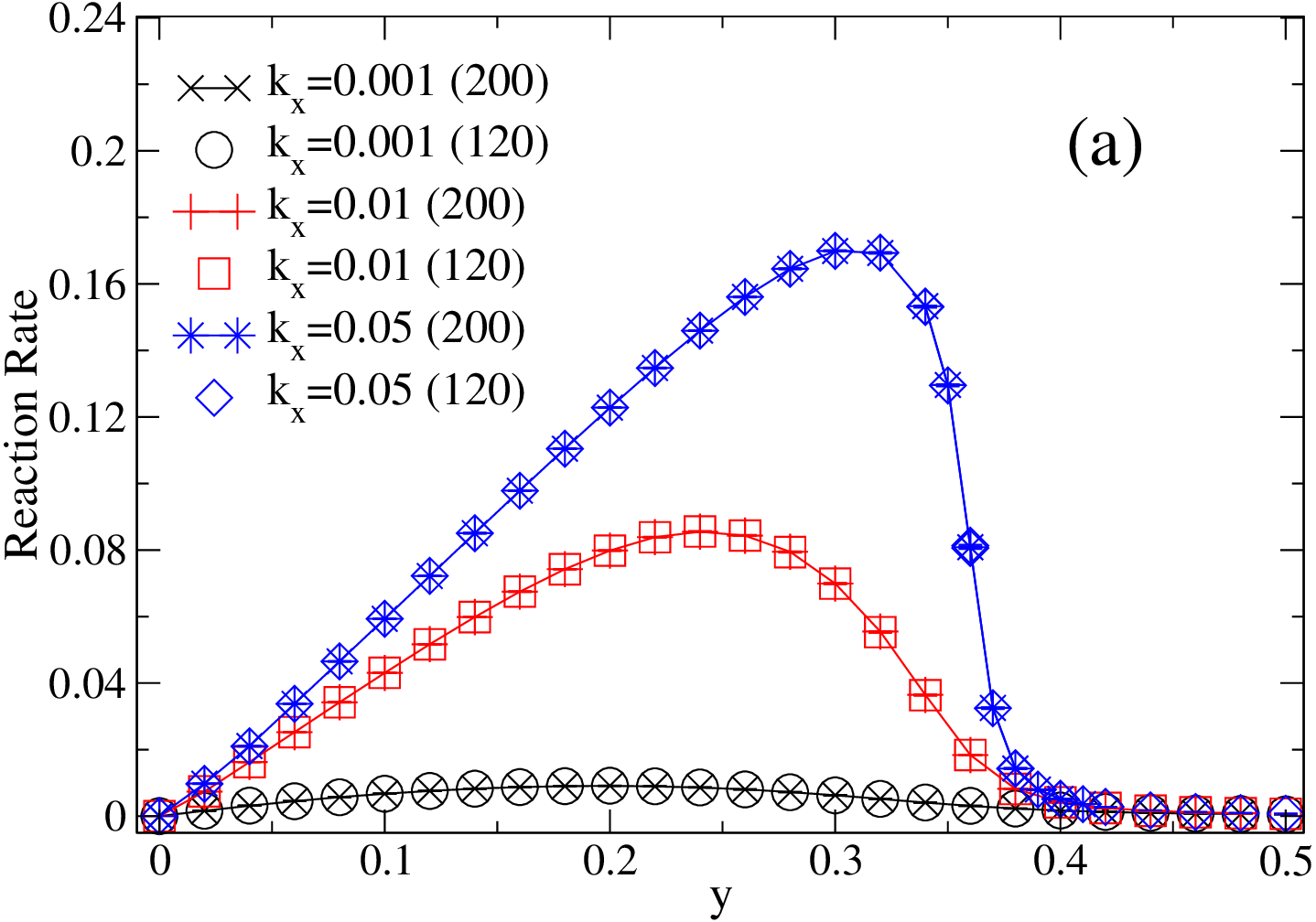}\\
\vskip 1.0truecm
\includegraphics[scale=.35]{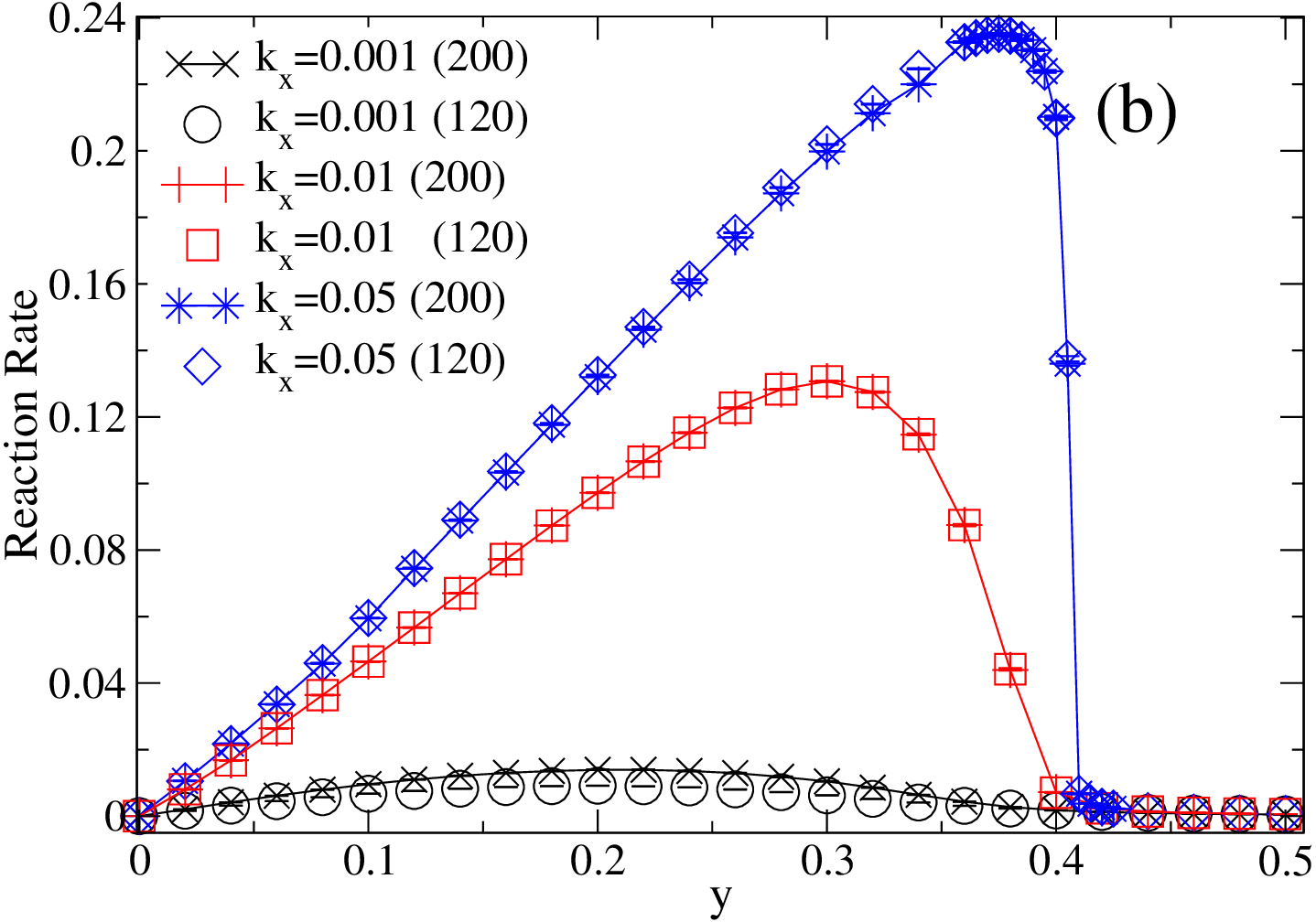}
\caption[]{
Reaction rates vs $y$ when the partial pressure of impurities is 
$y_{\rm x}=0.005$, 
(model B) with the values of $k_{\rm x}$ 
indicated in the figures. (a) 
$d=0$. (b) $d=0.9$. $k_{\rm co}$=$0.01$.
Additional data for a 200$\times$200 system show that finite-size effects 
are negligible. }
\label{f5}
\end{figure}

To further understand the effect of the CO diffusion, in Fig.~\ref{f5} 
we plot the reaction rates for the model with fixed $k_{\rm{co}}$ and 
different values of $k_{\rm{x}}$, when there is no diffusion ($d=0$), Fig.~\ref{f5}(a), 
and with a high diffusion attempt rate ($d=0.9$), Fig.~\ref{f5}(b).  Clearly, the diffusion term 
greatly increases the reaction rate of the system. Notice that  for the higher 
value of $d$, the range of $y$, for which the system produces CO$_2$, is slightly larger. 
For very small values of $k_{\rm{x}}$, the reaction rate is almost zero and 
almost totally insensitive to $d$. This is expected because, as we 
already mentioned, when the impurities 
do not desorb, or have a very low desorption rate, in the steady state the 
reaction rate is close 
to zero. The non-desorbing impurities take over the empty spaces, creating 
a barrier between the adsorbed O 
and CO (see Fig.~3 of Ref.~\cite{buen12}), such that they cannot react. Since there are no empty 
sites available, the rate of attempted diffusion moves has no effect in this case. 
Comparing Fig.~\ref{f5}(a) and (b), we again see
that the reaction rate increases with $k_{\rm{x}}$ and $d$.

\section{Conclusions}
\label{sec-Con}
In this work a modified version of the ZGB model for the reaction CO+O 
$\rightarrow$ CO$_2$ on 
a catalytic surface has been analyzed. With the aim to make the model richer and 
more amenable to present interesting nonequilibrium behaviors, and also to make it 
somewhat more realistic, we consider the simultaneous action of 
several processes that are present in real situations. 
We change the mechanism for the dissociative adsorption of O$_2$  
to incorporate 
experimental results that indicate that, once on the surface, the O atoms
experience a strong mutual repulsion. In our
model, the dissociative adsorption of O$_2$ requires 
two empty nnn sites and occurs only if the six nn of the entrance sites are not 
occupied by O. This mechanism is known as the eight-site rule 
~\cite{brun84,chan87,jame99}. 
The model also incorporates  CO desorption and diffusion that in real systems 
are related to temperature effects.  To further understand 
what happens in 
experimental conditions, particularly in industrial environments, we also study 
the effect of 
 impurities (X) in the gas phase. Once adsorbed on the surface, the 
impurities do not react with the other species and can only be desorbed.
Then, 
besides the variable $y$, proportional to the partial pressure of CO in the gas, that 
defines the ZGB model, our model has several other variables: $y_{\rm{x}}$ 
proportional to 
the partial pressure of X in the gas; $k_{\rm{co}}$ and $k_{\rm{x}}$, the 
desorption rates 
of CO and X, respectively; and $d$, the attempt rate for CO diffusion.  An evident
consequence of the modification of the entrance mechanism for oxygen is that the 
non-physical oxygen poisoned phase of the original ZGB model 
disappears~\cite{buen12, buen13}. Another consequence of the strong O 
repulsion is a considerable increase of the CO$_2$ production rate for 
values of $y<y_c$, where $y_c$ is the value of $y$ at which the system becomes 
filled mostly 
with CO. $y_c$ is smaller for the system with strong O repulsion, such 
that the range of values of $y$ for which the system remains reactive is significantly 
reduced for these systems.  We also 
find that the transition to the CO rich phase is smoother for the system 
with strong O repulsion, and the dependence of $y_c$ on the CO 
desorption rate is more accentuated: $y_c$ increases with increasing values of 
$k_{\rm co}$. Beyond a critical value of  $k_{\rm co}$, the discontinuous transition 
at $y_c$ is replaced by a smooth crossover.
These results can be explained by the fact that the 
strong repulsion between the O atoms, represented by the eight-site rule, hinders 
their entrance, thus favoring 
the entrance of the other species. If there are no impurities, the empty sites 
are filled 
with CO such that the CO poisoning occurs at lower values of $y$.  When 
impurities are added to the gas-mixture, the sites that cannot be filled with 
O can be filled with CO or X.  Since the 
adsorbed X do not react, they form a barrier between domains of CO and O 
impeding their reaction,
and unless the X desorption rate is high, their 
presence greatly diminishes the CO$_2$ production 
and eliminates the first-order transition between the productive and CO-poisoned phases
\cite{buen12,buen13,bust00,vill14}. 
In this scenario, where the 
strong O repulsion favors the existence of a larger number of empty sites and 
the adsorption of the other species, 
 the diffusion of CO also plays an important 
role in increasing the reaction rate. This study shows that CO 
desorption and diffusion can in certain ways counteract the negative effect of 
the presence of impurities on the reaction rates. This effect may 
explain why systems with impurities increase or recover their catalytic 
productivity with increased temperature. 

\section*{Acknowledgments}
G.M.B.\ is grateful for the hospitality of the Physics Department at Florida State University. 
P.A.R.\ acknowledges support by U.S. National Science Foundation Grant No. DMR-1104829.


\end{document}